\documentclass[prl,reprint,amsmath,amssymb,aps,superscriptaddress,y]{revtex4-2}

\usepackage{graphicx}
\usepackage{xcolor}
\usepackage{dcolumn}
\usepackage{soul}
\usepackage[normalem]{ulem}
\usepackage[T1]{fontenc}
\usepackage{bm}
\usepackage{amsmath}
\usepackage[caption=false]{subfig}
\usepackage{hyperref}

\newcommand{\ket}[1]{\lvert #1 \rangle}

\providecommand{\abs}[1]{\lvert#1\rvert}
\newcommand{\supi}{^{(i)}}

\usepackage{colortbl}
\usepackage{siunitx}
\usepackage{bbm}

\usepackage{color}
\definecolor{mygreen}{rgb}{0,0.5,0}
\definecolor{mygrey}{rgb}{0.5,0.5,0.5}
\definecolor{myred}{rgb}{0.75,0,0}
\definecolor{myblue}{rgb}{0,0,0.75}
\definecolor{mymagenta}{cmyk}{0,1,0,0.12}
\definecolor{mycyan}{cmyk}{1,0,0,0.12}
\definecolor{myorange}{rgb}{1,0.5,0}
\definecolor{myviolet}{rgb}{0.5,0.0,0.75}
\definecolor{mybrown}{rgb}{0.542969,0.269531, 0.0742188}

\newcommand{\ctext}[1]{{\color{cyan}#1}}
\newcommand{\rtext}[1]{{\color{myred}#1}}

\renewcommand{\ctext}[1]{{\color{black}#1}}
\renewcommand{\rtext}[1]{{\color{black}#1}}

\newcommand{\MSE}{\mathrm{MSE}}
\newcommand{\supSQL}{^{(\mathrm{SQL})}}
\newcommand{\subrot}{_\mathrm{rot}}

\newcommand{\downstate}{\left|\,\downarrow\,\right>}

\newcommand{\bJ}{\mathbf{J}}

\newcommand{\quadvec}{\bm{Q}}
\newcommand{\h}[1]{\hat{#1}}
\newcommand{\m}[1]{\langle #1 \rangle}
\newcommand{\estimator}[1]{\widehat{#1}}
\renewcommand{\estimator}[1]{\overset{*}{#1}}
\renewcommand{\estimator}[1]{{#1}_{\rm est}}

\usepackage[commentmarkup=uwave]{changes}

\definechangesauthor[color=myblue]{dbo}
\definechangesauthor[color=red]{jl}
\definechangesauthor[color=myviolet]{RJS}
\definechangesauthor[color=magenta]{MWM}

\begin{document}
\preprint{APS/123-QED}

\newcommand{\mytitle}{Improving Short-Term Stability in Optical Lattice Clocks by Quantum Nondemolition Measurement
}

\title{\mytitle}

\newcommand{\ICFO}{ICFO - Institut de Ci\`encies Fot\`oniques, The Barcelona Institute of Science and Technology, 08860 Castelldefels (Barcelona), Spain}
\newcommand{\ICREA}{ICREA - Instituci\'{o} Catalana de Recerca i Estudis Avan{\c{c}}ats, 08010 Barcelona, Spain}
\newcommand{\OP}{LNE--SYRTE, Observatoire de Paris, Universit\'{e} PSL, CNRS, Sorbonne Universit\'{e}, 61 avenue de l'Observatoire, F-75014 Paris, France}

\author{Daniel Benedicto Orenes}
\affiliation{\ICFO}

\author{Robert J. Sewell}
\affiliation{\ICFO}

\author{J\'{e}r\^{o}me Lodewyck}
\affiliation{\OP}

\author{Morgan W. Mitchell}
\affiliation{\ICFO}
\affiliation{\ICREA}

\begin{abstract}
We propose a multi-measurement estimation protocol for Quantum Nondemolition (QND) measurements in a Rabi clock interferometer. 
The method is well suited for current state-of-the-art optical lattice clocks with QND measurement capabilities. 
The protocol exploits the correlations between multiple nondestructive measurements of the initially prepared coherent spin state. 
A suitable Gaussian estimator for the clock laser detuning is presented, and an analytic expression for the sensitivity of the protocol is derived. 
We use this analytic expression to optimise the protocol using available experimental parameters, achieving an improvement of \SI{7.9}{dB} with respect to the standard quantum limit (SQL) in terms of clock stability. We also discuss the measurement back-action effects of our protocol into the atomic state. 
\end{abstract}

\maketitle

Optical lattice clocks (OLCs) use ultra-narrow (\SI{}{\milli\hertz}) transition lines in neutral atoms as absolute frequency references and have demonstrated fractional frequency instability below $\SI{E-18}{}$ \cite{hinkley_atomic_2013,OelkerNatPhot2019}, enabling new applications and fundamental studies in physics \cite{huntemann_improved_2014, lisdat_clock_2016,takamoto_test_2020, safronova_search_2018}. The Dick effect (DE), i.e., the sampling of the local oscillator frequency noise by the clock interrogation sequence, \rtext{remains a main contributor to the instability} of OLCs \cite{schulte_prospects_2020}. However, with the observation of second-scale atomic coherence times \cite{hinkley_atomic_2013,norcia_seconds-scale_2019}, and with the design of measurement sequences \cite{takamoto_frequency_2011,SchioppoNatPhot2017,OelkerNatPhot2019,NicholsonPRL2012,LodewyckPRA2009} able to evade the DE, OLC short-term instability, i.e., for integration time for which readout noise or DE contributes more than residual fluctuations of systematic effects, may soon be dominated by 
quantum projection noise (QPN). At that point, the instability of OLCs can be further reduced using techniques from quantum optics and cavity quantum electrodynamics (CQED) \cite{HuangThesis2019,polzik_entanglement_2016}.

Quantum nondemolition (QND) measurements of atomic spin \cite{SewellNP2013}, which produce spin squeezing \cite{takano_spin_2009, appel_mesoscopic_2009, sewell_magnetic_2012,HostenN2016}, can reduce QPN as part of back-action evading measurement protocols \cite{BraginskyFN1974, vasilakis_stroboscopic_2011,vasilakis_generation_2015,ColangeloN2017}, permit reuse of the spin ensemble to reduce dead time and the DE 
\cite{LodewyckPRA2009}, and are compatible with more sophisticated strategies to evade the DE using multiple ensembles and synchronous measurements  \cite{takamoto_frequency_2011,NicholsonPRL2012,SchioppoNatPhot2017,OelkerNatPhot2019} and improve phase locking of the local oscillator \cite{Shi12,kohlhaas2015phase,bowden_improving_2020}. 
Their efficacy is greatly enhanced using CQED methods \cite{Schleier-SmithPRL2010, BohnetNPhot2014} especially when optical lattice and cavity modes are structured to produce uniform coupling \cite{HostenN2016}. 
Existing and in-development OLCs incorporate optical cavities for cavity-enhanced
nondestructive measurement \cite{TaralloEFTF2017,BowdenSR2019,vallet_noise-immune_2017}, and multimode probing to achieve uniform coupling \cite{vallet_noise-immune_2017}, making  them very attractive systems for quantum noise reduction.

Paradigmatic QND-based enhancement protocols \cite{Kuzmich_1998, thomsen_continuous_2002,ChenPRA2014,appel_mesoscopic_2009} detect a population difference, e.g., the pseudo-spin component  $\h{J}_z = (\h{N}_\uparrow - \h{N}_\downarrow)/2$, where $\h{N}_\uparrow$ and $ \h{N}_\downarrow$ are clock transition level populations. Such protocols have been successfully implemented in microwave clocks \cite{HostenN2016,BohnetNPhot2014,Schleier-SmithPRL2010,kohlhaas2015phase,Lou10}, but are not directly applicable to the OLC scenario, where only the ground state population $\h{N}_\downarrow$ is readily available to dispersive measurement \cite{LodewyckPRA2009, vallet_noise-immune_2017,hobson_cavity-enhanced_2019} (see, however, Meiser et al. \cite{Meiser_2008} for a proposal for differential population QND via far-off-resonance probing). Spin-squeezing generated on a microwave transition has also been transferred coherently to an optical clock transition  \cite{pedrozo-penafiel_entanglement_2020}.

Here we propose a novel protocol that uses multiple QND measurements of the ground state population $\h{N}_\downarrow$  of the clock  transition to overcome the QPN and reduce the noise of laser detuning estimates obtained in Rabi spectroscopy.  The method is thus directly applicable to state-of-the-art OLCs as presently employed. We describe the quantum and classical statistics of atoms and probe light to second order, to obtain analytic results and  optimization under several scenarios within the Gaussian  approximation. 
For realistic QND detection parameters in current OLCs, we predict generation of squeezed states and a metrological improvement of \SI{7.9}{\decibel} relative to the best possible  frequency stability with projective measurements.

\begin{figure}
\centering
\begin{picture}(0,170)
\put(-125,0){\includegraphics[width= \columnwidth]{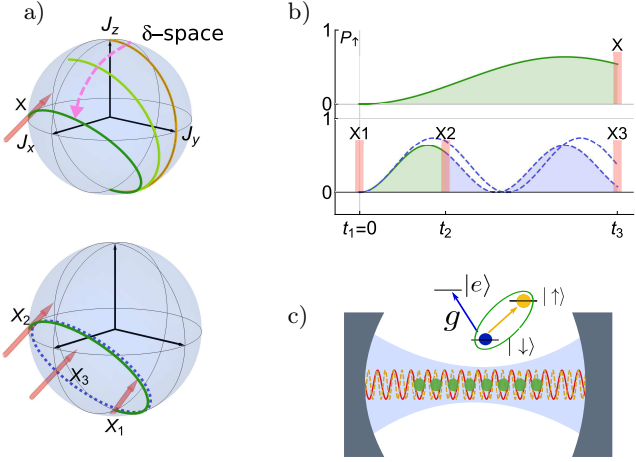}}
\put(-110,170){a)}
\put(-10,170){b)}
\put(-10,55){c)}
\end{picture}
\caption{
a) Bloch-sphere representation of protocols: upper panel shows reference Rabi sequence used in state-of-the-art OLCs, with a single measurement (labeled $X$) at the end of the sequence, used to define the SQL.
The colored lines correspond to different detunings, with the dark green line corresponding to the detuning maximizing the sensitivity of the clock, and the red arrows indicate when the measurement is made.
Lower panel shows protocol with three QND measurements. 
The sensitivity of this protocol is shown as the red curve in \autoref{fig:sensitivity}.
b) Upper/lower panels show excited state population $P_{\uparrow}$ of typical Rabi/three-measurement protocol (green/blue) as a function of time. Red bars indicate the timing of the QND measurements. 
The Rabi frequency and detuning have been adjusted such that the total interrogation time remains the same as in both protocols. The two dashed blue lines in the lower panel correspond to slightly different detunings to illustrate the sensitivity of the protocol to the detuning of the clock laser.
c) Pictorial representation of the experimental situation, showing trapping light (red), clock laser light (yellow) for inducing Rabi oscillations between $\ket{\downarrow}^{\otimes N} \leftrightarrow \ket{\uparrow}^{\otimes N}$, and probing light (light blue) coupling $\ket{\downarrow}, \ket{e}$ to nondestructively measure the ground state population. 
The generic representation of the energy levels relevant for this problem is also shown, typical in the Alkaline-earth-like atoms such as Sr or Yb used in OLCs. 
\label{fig:ProtocolBloch} }
\end{figure}

\paragraph{Definition of the problem.}
As illustrated in \autoref{fig:ProtocolBloch}, we consider an ensemble of $N$ three-level systems, in which $\left|\downarrow\right>$ and $\left|\uparrow\right>$ are the ground and excited states of the clock transition, respectively.
To understand the quantum noise in this scenario, we describe the spin system by the  collective spin operators $\h{J}_\alpha = \sum_{i=1}^N \h{\jmath}_\alpha\supi$, where $\boldsymbol{\h{\jmath}}^{(i)}$ is the vector pseudo-spin operator describing atom $i$ and $\h{\jmath}_0\supi/\hbar = \mathbbm{1}/2$ \footnote{From now and throughout the rest of the work, we set $\hbar =1$. Vectors are written in bold characters, while components of vectors use normal italic notation. Mean field values of operators $\h{\mathcal{O}}$ are indicated using bracket $\langle \mathcal{O} \rangle$ notation.}.
Rabi oscillations on the atomic transition driven by a clock laser operating as a local oscillator are governed by the rotating-frame Hamiltonian \ctext{$\h{H}_{\rm rot} = \Omega \h{J}_x - \delta \h{J}_z$}, where $\delta = \omega_L - \omega_0$ is the detuning of the clock laser with respect to the atomic transition and $\Omega$ is the corresponding Rabi frequency. 

QND measurements are described by introducing a ``meter'' in the form of a pulse of light with mode operator $a$ and thus field quadratures $\h{X} \equiv (\h{a}^{\dagger} + \h{a})/2$ and $ \h{P} \equiv i( \h{a}^{\dagger} - \h{a})/2$, such that $[\h{X},\h{P}] = i/2$. During a measurement, the meter interacts with the atoms via Hamiltonian $\h{H}_{\rm qnd}$ for time $\tau$, such that \ctext{$\tau \h{H}_{\rm qnd} = g \h{n} \h{N}_{\downarrow} \approx g \abs{\alpha}^2 \h{N}_{\downarrow}  +  2 g \abs{\alpha} \varDelta \h{P} \h{N}_{\downarrow}$} \rtext{\cite{SI2}}, where $g$ is the atom-light coupling constant \footnote{$g = \frac{1}{S_{\rm eff}} \frac{3 \lambda^2 }{2 \pi} \frac{\Delta_{\rm qnd}/\Gamma}{s+4 (\Delta_{\rm qnd}/\Gamma)^2}$, with $s$ a saturation parameter that depends on the intracavity power and $S_{\rm eff}$ an effective geometric coupling parameter. In this formula $\Delta_{\rm qnd}$ is the probe light detuning with respect to the atomic transition. Realistic experimental parameters have been chosen according to \cite{vallet_noise-immune_2017}.}, 
$\h{n} = \h{a}^{\dagger} \h{a}$ is the photon number operator, and $\varDelta \h{A} \equiv \h{A} - \langle A \rangle$ indicates a deviation from the mean. 
The \ctext{approximation} follows assuming that the input pulses are coherent states, with amplitude $\alpha = i \abs{\alpha}$ and thus $\langle X \rangle = 0$, $\langle P \rangle = \abs{\alpha}$, $\langle n \rangle = \abs{\alpha}^2$. We assume here that the atoms are uniformly coupled to the cavity field.
The first term, which commutes with the optical variables, represents a light-shift term which induces a rotation of the atomic state about $J_z$, but does not contribute to the optical signal,
and can be neglected \footnote{The effect of this term \rtext{can} be cancelled by precise control of the optical power and coupling constant \rtext{i.e., by making $g \vert \alpha \vert^2 = 2 \pi j, j \in \mathbbm{Z}$} \cite{thomsen_continuous_2002}, or by adapting more sophisticated optical probing techniques \cite{saffman_spin_2009} }.

The term $2 g \abs{\alpha} \varDelta \h{P} \h{N}_{\downarrow}$ in $\tau \hat{H}_{\rm qnd}$ contributes a phase shift $g \m{N_\downarrow}$ to the probing light.
We also assume that $g \m{N_\downarrow} \ll \pi$, which can always be satisfied by choice of probe light detuning, and that the QND pulse duration is short compared with the Rabi dynamics, $\tau \ll \pi/\Omega$. Then the 
postinteraction value $\h{X}^{\rm (out)}=\h{X}^{\rm (in)} + g \abs{\alpha} \h{N}_{\downarrow}$
serves as a linear ``pointer'' to indicate the ground state population $\h{N}_{\downarrow}=\h{J}_0 - \h{J}_z$
at the time $t$ when the pulse interacts with the atoms.
A strong homodyne readout of $\h{X}^{\rm (out)}$ completes the QND measurement.

\newcommand{\bmu}{\boldsymbol{\mu}}

\paragraph{Measurement protocol and estimation strategy.}
We assume an ensemble of $N$ atoms is prepared by optical pumping in the ground state $\downstate^{\otimes N}$ of the clock transition, and allow for fluctuations in the atom number due to the random processes involved in loading the trapped atoms.
The clock laser drives the transition causing the spin state to execute Rabi oscillations with frequency $\Omega$. 
A set of QND measurements of $\h{X}^{\rm (out)}$ is performed at times $t_l$, with corresponding outcomes ${\bf X} \equiv \{X_l\}$, using a number of photons $\abs{\alpha_l}^2$ in each measurement. 
We use ${\bf X}$ to build an estimator $\estimator{\delta}$ for the clock laser detuning(see Supplemental Material ~\cite{SI2}). 
For large atom number $N$, ${\bf X}$ can be well approximated by a Gaussian random variable with distribution 
\begin{equation}
    P(\textbf{X}|\delta) = \frac{1}{\sqrt{(2 \pi)^d|\Gamma_X|}}\exp[ -\frac{1}{2}(\textbf{X}-\bmu)^T \Gamma_X^{-1} (\textbf{X}-\bmu)]
\end{equation}
where $d$ is the length of ${\bf X}$, $\bmu$ and $\Gamma_X$ are the mean and covariance matrix of ${\bf X}$, respectively, which are functions of $\delta$. After linearization about $\delta_0$, the nominal value of $\delta$, and defining $\bmu_0 \equiv \bmu(\delta_0)$, the maximum likelihood estimator for $\delta$ is \cite{SI2}
\begin{equation}
\label{eq:DeltaEstimator}
    \estimator{\delta} = \delta_0 + \frac{(\bmu')^T \Gamma_X^{-1} (\textbf{X}-\bmu_0)}{(\bmu')^T \Gamma_X^{-1}\bmu'}
\end{equation}
where $\bmu' \equiv \partial_{\delta} \bmu|_{\delta_0}$. 
The sensitivity of the protocol to the laser detuning $\delta$ is given by the mean squared error (MSE) of the detuning \footnote{In case of maximum likelihood, unbiased estimators, the MSE is equivalent to the estimator variance. The proof that our estimator is unbiased can be found in the supplementary material}:
\begin{equation}
\label{eq:DeltaMSE}
    {\rm MSE (\delta)} \equiv E[(\estimator{\delta} - \delta)^2] = \frac{1}{(\bmu')^T ~ \Gamma_X^{-1} ~ \bmu'}.
\end{equation}

To compute $\bmu$ and $\Gamma_X$ we employ established covariance matrix methods \cite{MadsenPRA2004, KoschorreckJPAMOP2009, Colangelo2013,kraus_entanglement_2003,molmer_estimation_2004,PetersonPRA2006}. 
We describe the spin system and optical probe pulses with the phase-space vector $\h{\bm{V}} =  \h{\bm{J}} \oplus \left(\bigoplus_{i=1}^{N_{\rm pulses}} \varDelta \h{\quadvec}_i \right)$, where $\h{\bm{J}}=(\h{J}_0,\h{J}_x,\h{J}_y,\h{J}_z)^T$, and $\h{\quadvec}_i \equiv ( \h{X}_i, \h{P}_i)^T$ describes the $i$th optical pulse. 
We assume that the state is Gaussian, and remains so during the entire measurement sequence.
The system is thus completely characterized by the vector of first moments $\langle \bm{V} \rangle$ and the covariance matrix $\Gamma_V = 1/2 ~ \langle \bm{V} \rtext{\otimes} \bm{V} + (\bm{V} \rtext{\otimes} \bm{V})^{T}\rangle - \langle \bm{V} \rangle \rtext{\otimes} \langle \bm{V} \rangle$.
$\langle \bm{V} \rangle$ and $\Gamma_V$ evolve deterministically through the sequence of coherent Rabi oscillations generated by $\h{H}_{\rm rot}$, and sudden
light-matter interactions generated by $\h{H}_{\rm qnd}$. 
The transformation of the state is computed by integrating  $d\h{\mathbf{V}}/dt = {-i} [\h{\mathbf{V}},\h{H}_\mathrm{qnd}]$ and dropping terms beyond first order in quantum fluctuations, to find linear input-output relations.
We include the effect of loss and decoherence due to atom-photon scattering during the probing parametrized by $\eta = \eta_{\gamma} \abs{\alpha}^2$, the fraction of atoms that scatter a photon due to the interaction with a probe pulse containing $\abs{\alpha}^2$ photons with incoherent scattering rate $\eta_\gamma$ \rtext{\cite{SI2}}. \rtext{Multi-pulse sequences are constructed analogously.} The model gives unsightly but useful analytic results for $\bmu$ and $\Gamma_X$, which can be read off directly from $\langle \bm{V} \rangle$ and $\Gamma_V$ \cite{SI2,KoschorreckJPAMOP2009,Colangelo2013}.

\begin{figure}[ht]
\begin{picture}(0,120)
	\put(-125,0){\includegraphics[width=.24\textwidth]{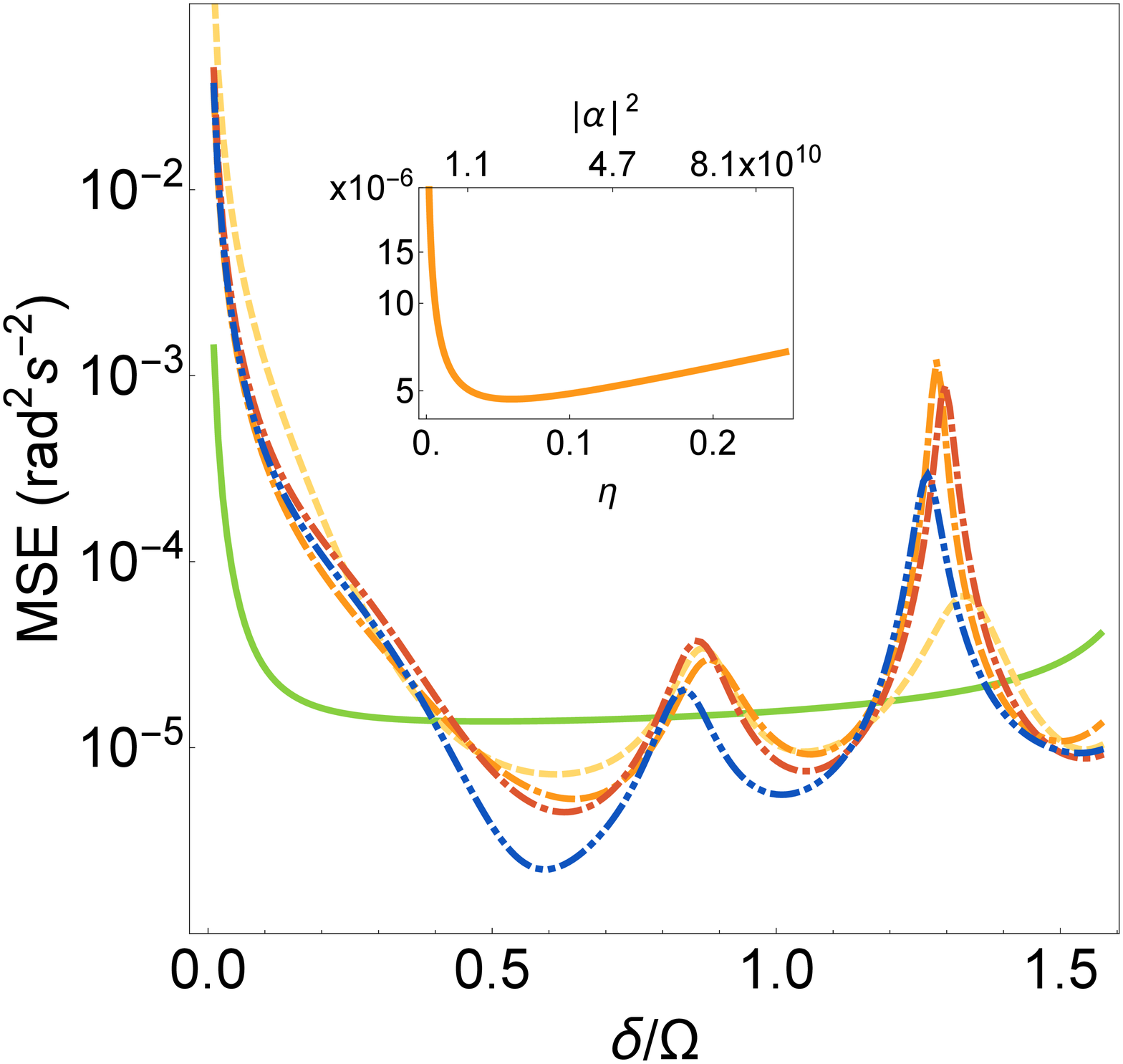}}
	\put(-100,105){a)}
	\put(0,0){\includegraphics[width=.24\textwidth]{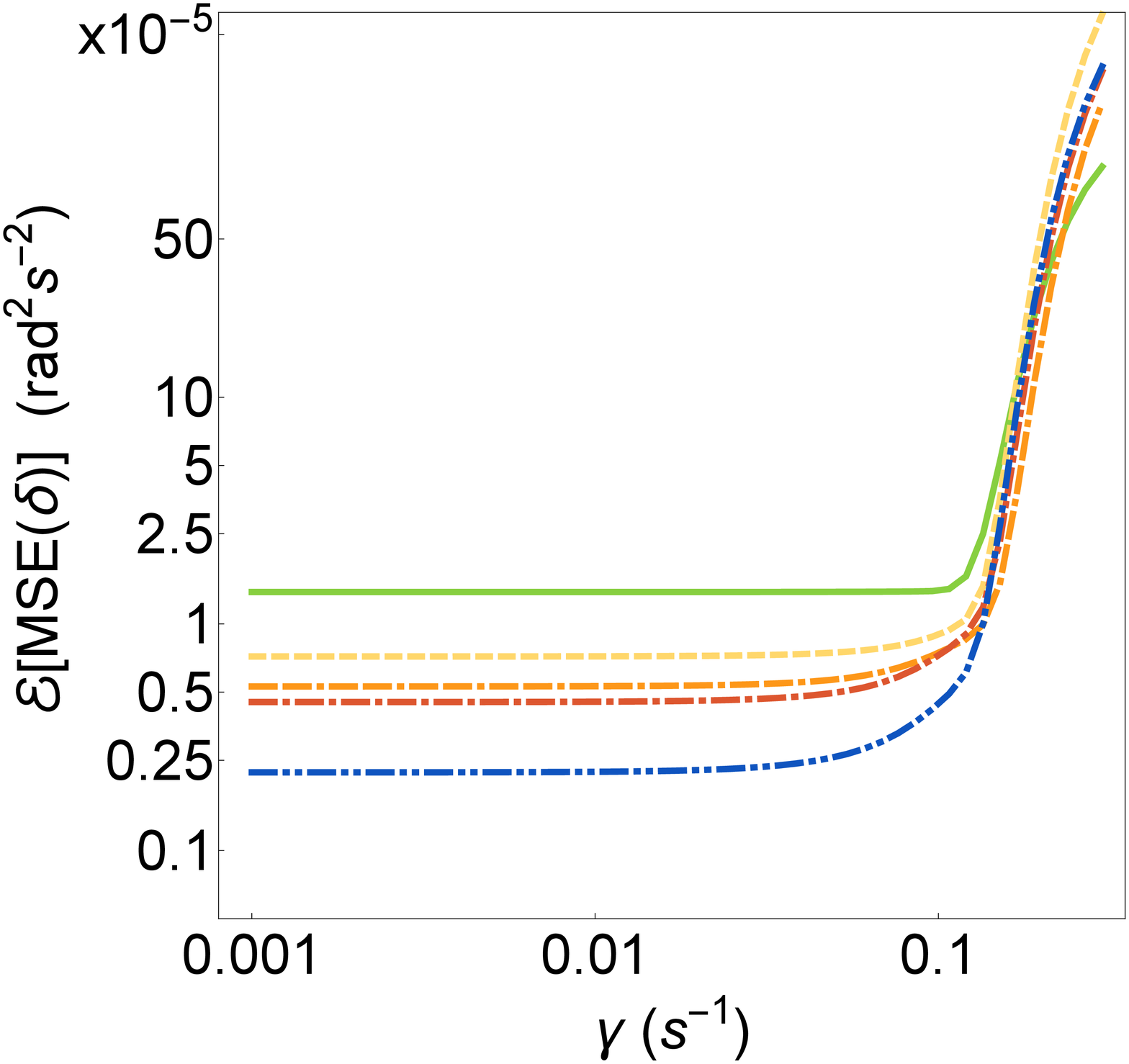}}
	\put(25,105){b)}
\end{picture}
\caption{
a) Sensitivity of the QND measurement protocol.  
Curves show the MSE of the detuning $\delta$ as a function of the drive laser detuning $\delta$, as per \autoref{eq:DeltaMSE}.
Atom number is $N = \SI{2 e4}{}$, atom-light coupling $g \simeq \SI{3.05 e-7}{rad/atom}$ obtained for an intracavity power of $\approx \SI{4}{mW}$.
The solid green curve shows the results for the reference protocol described in the main text, defining the standard quantum limit. 
The dashed yellow curve shows the three-measurement protocol with $t_1 =\pi/\Omega$, $t_2 = 3\pi/\Omega$ (\emph{a priori} values) and  $\abs{\alpha_1}^2 = \abs{\alpha_2}^2 = \abs{\alpha_3}^2 = \abs{\alpha_{\rm ref}}^2 = \SI{7.18 e9}{}$ (found by minimization of MSE with respect to $\abs{\alpha_{\rm ref}}^2$). 
The dashed-dashed-dot orange curve shows the three-measurement protocol with measurement timings $t_1 = 0$, $t_2 \approx 1.79 \pi/\Omega$ and $t_3  \approx 3 \pi/\Omega$, (found by optimization of MSE with respect to $t_l$) and keeping the values for $\abs{\alpha_1}^2 = \abs{\alpha_2}^2 = \abs{\alpha_3}^2 = \abs{\alpha_{\rm ref}}^2 = \SI{7.14 e9}{}$.
The dot-dashed red curve shows the three-measurement protocol found by optimization of the timings and the number of photons per measurement pulse while keeping the total equal to the previous protocol i.e. $\abs{\alpha_1}^2 + \abs{\alpha_2}^2 + \abs{\alpha_3}^2 \leq 3 \abs{\alpha_{\rm ref}}^2$. In this case, $t_2 \approx 1.27 \pi/\Omega, t_3 \approx 3 \pi/\Omega$, and $ \abs{\alpha_1}^2 = \SI{3.37e9}{} ,  \abs{\alpha_2}^2 = \SI{5.4e9}{},  \abs{\alpha_3}^2  = \SI{1.28 e10}{} $.
The dot-dot-dashed blue curve shows a full optimization of the timings and number of photons without the previous constraint, with optimum parameters $t_2 \approx 1.29 \pi/\Omega, t_3 \approx 3 \pi/\Omega$, and $ \abs{\alpha_1}^2 = \SI{9.77e9}{} ,  \abs{\alpha_2}^2 = \SI{2.12e10}{},  \abs{\alpha_3}^2  = \SI{1.35e11}{} $.
The inset shows the MSE with the timings corresponding to the orange protocol (i.e., same measurement timing, and equal measurement pulses) at the optimum detuning $\delta_0/\rtext{\Omega} \simeq 0.6$ for varying measurement strength (photon number for each measurement pulse) parametrized by the fraction of atoms $\eta$ that suffers incoherent scattering during the protocol. 
b) Numerical results of the averaged single shot estimator variance $\mathcal{E}[\MSE(\delta)]$ as defined in the main text as a function of the clock laser phase noise parameter $\gamma$, with the same color coding as in panel a).}
\label{fig:sensitivity}\label{fig:allan}
\end{figure}

\paragraph{Reference protocol.}
The canonical Rabi sequence starts with exactly $N$ atoms in the state $\downstate^{\otimes N}$, which evolve under $\h{H}\subrot$ for a time $t$, after which a projective measurement of $\h{N}_{\downarrow}$ (or  equivalently $\h{N}_{\uparrow}$ or $\h{N}_{\uparrow} - \h{N}_{\downarrow}$) is made. 
We take the precision obtained by this sequence to define the standard quantum limit (SQL) of a \rtext{Rabi-spectroscopy} OLC. 
An equivalent protocol can be implemented using our formalism as follows: the initial state $\downstate^{\otimes N}$ implies $\langle \bJ \rangle = (1,0,0,-1) \langle N \rangle/2$, with $\Gamma_J = \mathrm{diag}(0,1,1,0)\langle N \rangle/2$. 
The evolution under $\h{H}\subrot$ induces a $\delta$-parametrized O(3) rotation (the $\h{J}_0$ component is unchanged) of $\langle \bJ \rangle$ and $\Gamma_J$, while the measurement is of $\h{J}_0-\h{J}_z$. The resulting MSE from \autoref{eq:DeltaMSE} is $\MSE\supSQL = \kappa_1 N^{-1} + \kappa_2(g \abs{\alpha}^2 N)^{-1}$, where $\kappa_1,\kappa_2$ are constants \rtext{that depend on} the chosen sequence parameters $\Omega, \delta$.

\paragraph{Three-pulse protocol.}
The new protocol includes three QND measurement pulses. The ensemble of atoms is prepared in the ground state $\downstate^{\otimes N}$, and we allow for Poisson-distributed atom number fluctuations in the input state \cite{ColangeloPRL2017}.
The state is then allowed to evolve under $\h{H}_{\rm rot}$ and subject to QND measurements at times $t_1=0$, i.e., immediately after state preparation, $t_2$ and $t_3$, with number of photons per measurement pulse $\abs{\alpha_1}^2,\abs{\alpha_2}^2,\abs{\alpha_3}^2$. In essence, the first QND pulse is used to calibrate the number of atoms, while the other two act as squeezing and read-out pulses.
The $\delta-$dependent $\MSE$ is calculated using \autoref{eq:DeltaMSE}, optimized with respect to measurement times $t_i$ and photon numbers $\abs{\alpha_i}^2$.

\paragraph{Sensitivity optimization and clock stability.}
\autoref{fig:sensitivity}(a) shows the sensitivity of our reference protocol as a function of the driving laser detuning $\delta$ using different parameters for a  typical atom number $N = \SI{2 e4}{}$.
Each dashed line corresponds to a different choice of  settings $\{ \abs{\alpha_l}^2, t_l\}$, found by minimizing $\rm MSE(\delta)$ with respect to $\delta$ and $\{\abs{\alpha_l}^2, t_l\}$, subject to constraints on $\{\abs{\alpha_l}^2, t_l\}$, e.g., $\sum_l \abs{\alpha_l}^2 = \abs{\alpha}^2_{\rm tot}$ (see caption in \autoref{fig:sensitivity}(a)).
With \emph{a priori} timing of the measurements and optimized photon numbers, we observe an improvement in the sensitivity of $\sim\,$\SI{2.8}{dB} with respect to the SQL. 
With further optimization of the pulse timings an improvement of $\sim\,$\SI{4.2}{dB} relative to the SQL is achieved, and when both timing and measurement strength (i.e. photon number) are fully optimized, the protocol achieves $\sim\,$\SI{7.9}{dB} improvement over the SQL.

\paragraph{Allan variance.} 
Current state-of-the-art OLCs are limited by the DE rather than the QPN. Nevertheless, there are situations where the enhanced sensitivity of the proposed protocol will be advantageous. For instance, synchronous clock comparison measurements allow one to access the QPN-limited precision, and are important in precision measurements for fundamental tests, accurate measurements of fundamental constants, geodesy, or searches of exotic physics \cite{lange_improved_2021,godun_frequency_2014,takamoto_test_2020,safronova_search_2018,takano_geopotential_2016}. In addition, the functional relation of the sensitivity with respect to the detuning in our protocol differs from that of a standard clock protocol. In all these situations, it becomes important to assess the quality of our estimation taking into account the random phase fluctuations of the driving clock laser. The corresponding Allan deviation of the fractional frequency instability of an OLC using our protocol can be written as $\sigma^2(\tau_{\rm av}) = \mathcal{E}[\MSE(\delta)]T_c/(\omega_0^2 \tau_{\rm av})$ \cite{allan_statistics_1966,pezze_heisenberg-limited_2020}, where $T_c$ is the cycling time of the clock, and $\tau_{\rm av}$ is the averaging time, $\omega_0$ is the atomic clock transition frequency, and
\begin{equation}
    \mathcal{E}[\MSE(\delta)] = \int_{\estimator{\delta} - \varepsilon}^{\estimator{\delta} + \varepsilon} d\delta ~ P_{\gamma}(\delta) \MSE(\delta),
    \label{eq:av_est}
\end{equation}
where $P_{\gamma}(\delta) = 1/(\gamma \sqrt{2 \pi}) \exp[-(\estimator{\delta}-\delta)^2/2 \gamma^2]$ is the probability density for the random variable $\delta$ for a clock laser with a one sided power spectral density $S(\nu) = 2 \gamma^2/\nu$, which models current state-of-the-art clock lasers in which the dominant source of noise is flicker noise \cite{braverman_impact_2018,andre2004stability}. \rtext{To avoid including unphysical divergences in the integral, we take $\varepsilon = \pi\Omega$}~\footnote{\rtext{Due to the linearization leading to \autoref{eq:DeltaEstimator}, $\rm{MSE}(\delta)$, valid near $\delta = \delta_{\rm est}$, diverges near $\delta=\delta_0$. As a check on the approach of limiting the integral to $\delta_\mathrm{est} \pm \varepsilon$, we follow \cite{braverman_impact_2018} and integrate over all $\delta$ the regularized $\widetilde{\rm{MSE}}(\delta) \equiv \min[\rm{MSE}(\delta), \SI{4}{\radian\squared\per\second\squared}]$, where $\SI{4}{\radian\squared\per\second\squared}$ is an upper bound on the uncertainty in light of the linewidth. The results are consistent with the simpler method used here.}}. \ctext{The expression for $\sigma^2(\tau_{\rm av})$ does not include the effect of dead time in the clock cycle.} \autoref{fig:allan}(b) shows the results of computing the averaged single shot estimator variance $\mathcal{E}[\MSE(\delta)]$ as a function of the laser noise parameter $\gamma$ for the different curves in \autoref{fig:sensitivity}(a).
The different configurations of the three-measurement protocol show consistent improvement with respect to the reference protocol for a wide range of laser noise parameters $\gamma < 0.1$ \cite{SI2}, within current experimental capabilities of clock lasers \cite{schulte_prospects_2020}.

\paragraph{Measurement back action \& spin squeezing.}
Entanglement among the atoms can be detected using spin-squeezing inequalities \cite{toth_optimal_2007} that constrain the statistics of $\h{\boldsymbol{J}}$. In particular, if  $\{\h{J}_j,\h{J}_k,\h{J}_l\}$ are orthogonal components of $\h{\boldsymbol{J}}$, all separable states with uncertain particle number \ctext{obey several spin squeezing inequalities~\cite{hyllus_entanglement_2012,SI2}}. In the large-$N$ scenario, we can define a witness \cite{SI2}:
\begin{equation}
    \xi^2 \rtext{\equiv  \min_{\{\h{\jmath},\h{k},\h{l}\}}} \frac{\langle N \rangle (\Delta \h{J}_l)^2} { \langle \h{J}_j \rangle^2 + \langle \h{J}_k \rangle^2 }
    \label{eq:witness}
\end{equation} 
\rtext{where $\{\h{\jmath},\h{k},\h{l}\}$ are orthonormal}.  $\xi^2<1$ indicates the presence of spin squeezing and entanglement. 

\begin{figure}[t]
\begin{picture}(0,120)
	\put(-125,0){\includegraphics[width=.24\textwidth]{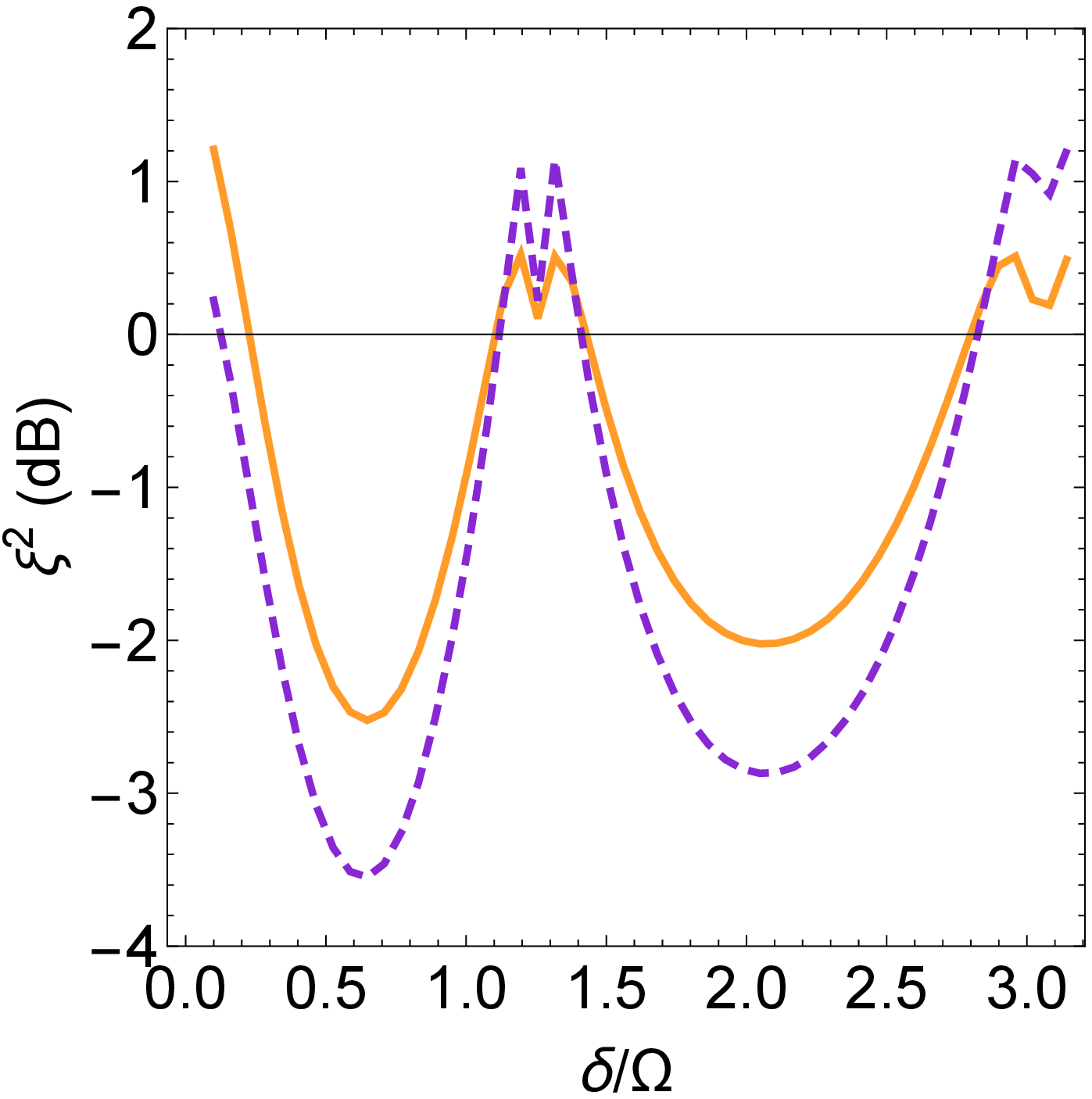}}
	\put(-100,105){a)}
	\put(0,0){\includegraphics[width=.24\textwidth]{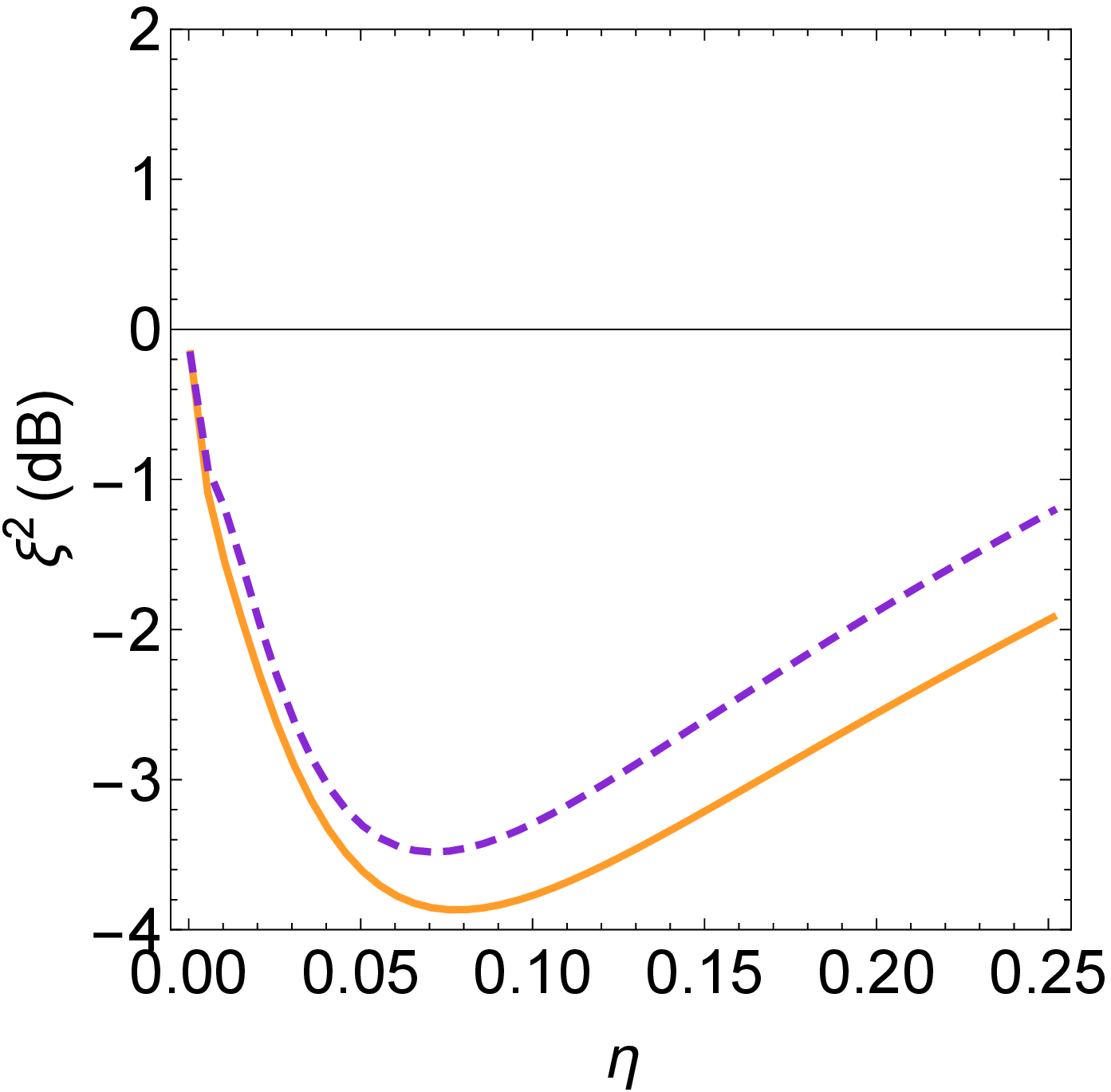}}
	\put(25,105){b)}
	\end{picture}
\caption{Numerical evaluation of the spin-squeezing witness $\xi^2$ in \autoref{eq:witness} using the three-measurement protocol with equal number of photons per measurement, as in the dot-dashed orange curve of \autoref{fig:sensitivity}.
We plot $\xi^2$ as a function of (a) the clock laser detuning $\delta$ for the same protocol with two different numbers of photons per measurement $\abs{\alpha}^2 = \{\SI{7.18e9}{} , 2 \times \SI{7.18e9}{}\}$ and (b) the measurement strength for two different detunings $\delta/\Omega = \{0.642, 0.5\}$.
In both plots the orange line corresponds to the parameters giving the optimized clock sensitivity, and the dashed purple curve shows the same protocol with different parameters. 
\label{fig:squeezing}}
\end{figure}

In \autoref{fig:squeezing} we plot $\xi^2$  immediately prior to the third QND measurement of the three-measurement protocol using the same number of photons per measurement (as in the orange line in \autoref{fig:sensitivity}) as a function of the clock laser detuning and  measurement strength. 
While the qualitative behavior of the entanglement witness is similar to that the clock sensitivity, we note that it is not a reliable indicator of the optimum sensitivity. In \autoref{fig:squeezing}.a we can see how choosing a different measurement strength leads to larger squeezing of the atomic state, while this configuration is not the optimum for the protocol sensitivity.
This is not unexpected, since the witness does not capture correlations due to the coupling of $\h{J}_0$ into the QND measurement. 
Using a spin squeezing witness which could account for the correlations between $\h{J}_0$ and the other spin variables would be desirable. However, to the best of our knowledge, such witness has not been presented in the literature.

\paragraph{Conclusions and outlook.}
In this Letter, we propose and analyze a minimal three pulse QND-based measurement protocol directly applicable to state-of-the-art OLCs using Rabi protocols and limited by QPN.
By exploiting a multimeasurement estimation strategy and measurement induced correlations, we show a \SI{7.9}{dB} enhanced single shot sensitivity of the clock with respect to the SQL for $N = \SI{2 e4}{}$ atoms. 
Our calculation techniques allow for straight-forward optimization of relevant experimental parameters, such as the timing and strength of each QND measurement. Using an entanglement witness derived from spin-squeezing inequalities for fluctuating atom number, we infer that the protocol generates entanglement.

The proposed protocol can improve the \rtext{short-term} stability of clock experiments that include strategies to evade the DE using multiple ensembles and synchronous measurement \cite{takamoto_frequency_2011,NicholsonPRL2012,SchioppoNatPhot2017,OelkerNatPhot2019}, and could be combined with spin-echo protocols to counteract inhomogeneous broadening induced by nonuniform atom-light coupling \cite{bowden_improving_2020}. Possible extensions of the framework presented here can extend the Gaussian description of the system to include diffusion terms \cite{serafini}, naturally incorporating into the results the noise characteristics of the clock laser. \rtext{We leave for future work the question of whether similar protocols with additional measurements, continuous probing and/or controlled phase shifts in the drive laser could give further enhancement.}

This work was supported by the European Union from the European Metrology Programme for Innovation
and Research (EMPIR) Project No. 17FUN03-USOQS. EMPIR projects are cofunded by the European Union's Horizon 2020 research and innovation program and the EMPIR participating states. We also acknowledge H2020 QuantERA ERA-NET Cofund Q-CLOCKS (PCI2018-092973 project funded by MCIN/ AEI /10.13039/501100011033/ FEDER “A way to make Europe”), H2020 FET Quantum Technologies Flagship project MACQSIMAL (Grant No. 820393), the Spanish Ministry of Science projects OCARINA (PGC2018-097056-B-I00 project funded by MCIN/ AEI /10.13039/501100011033/ FEDER “A way to make Europe”) and ``Severo Ochoa'' Center of Excellence CEX2019-000910-S. Generalitat de Catalunya: CERCA, AGAUR Grant No. 2017-SGR-1354, Secretaria d'Universitats i Recerca (project QuantumCat, 
Ref. No.~001-P-001644), Fundaci\'{o} Privada Cellex, and Fundaci\'{o} Mir-Puig.

\bibliographystyle{apsrev4-1no-url}
\bibliography{OLC_dbo_good}

\end{document}


\preprint{APS/123-QED}

\newcommand{\mytitle}{Supplementary Materials for ``Improving short-term stability in optical lattice clocks by quantum non-demolition measurement"}

\title{\mytitle}

\newcommand{\ICFO}{ICFO - Institut de Ci\`encies Fot\`oniques, The Barcelona Institute of Science and Technology, 08860 Castelldefels (Barcelona), Spain}
\newcommand{\ICREA}{ICREA - Instituci\'{o} Catalana de Recerca i Estudis Avan{\c{c}}ats, 08010 Barcelona, Spain}
\newcommand{\OP}{LNE--SYRTE, Observatoire de Paris, Universit\'{e} PSL, CNRS, Sorbonne Universit\'{e}, 61 avenue de l'Observatoire, F-75014 Paris, France}

\author{D.~Benedicto Orenes}
\affiliation{\ICFO}

\author{Robert J. Sewell}
\affiliation{\ICFO}

\author{J\'{e}r\^{o}me Lodewyck}
\affiliation{\OP}

\author{Morgan W. Mitchell}
\affiliation{\ICFO}
\affiliation{\ICREA}

\maketitle

\tableofcontents

\section{Atom-light coupling Hamiltonian and interaction matrix}

The atom-light coupling Hamiltonian is
\begin{equation}
    \h{H}_{\rm qnd} = \frac{g}{\tau} \h{N}_{\downarrow} \h{n} = \frac{g}{\tau} (\h{J}_0 - \h{J}_z) \h{a}^{\dagger} \h{a},
\end{equation}
and acts for a time $\tau$.  Here $\h{J}_i$ are collective spin operators as in the main text, $\h{a},\h{a}^{\dagger}$ are the annihilation/creation operators of the light field such that $[\h{a},\h{a}^\dagger] = 1$, and the photon number operator is $\h{n} = \h{a}^{\dagger} \h{a}$. 
This Hamiltonian produces a phase shift $\phi \equiv g\h{N}_\downarrow$ of the optical field. With a large optical detuning and large $\h{n}$, it is possible to ensure both $\phi \ll \pi$ and a high resolution in the measurement of $\h{J}_z$ \cite{KoschorreckJPAMOP2009, KoschorreckPRL2010a}.  Simultaneously, $\h{H}_{\rm qnd}$ produces a rotation of $\h{\mathbf{J}}$ about the $z$ axis by an angle $\theta \equiv g \h{n}$.  In the single-probe scenario considered here, $\theta$ is not in general small. One can, however, ensure that $\theta$ is within a small angle of an integer multiple of $2\pi$, leaving $\h{\mathbf{J}}$ nearly unchanged apart from a small residual rotation about $z$ due to fluctuations in $\h{n}$. We note that multi-frequency probing methods \cite{LodewyckPRA2009, hobson_cavity-enhanced_2019} can in principle null also the mean value of $\theta$, but for this work we consider only the simpler scenario of a single probe frequency.

We define field quadrature operators  $\h{X}= (\h{a}^{\dagger} + \h{a})/2$, $\h{P} = i( \h{a}^{\dagger} - \h{a})/2$ such that $[\h{X},\h{P}] = i/2$. For a coherent state $\ket{\alpha}$, the mean photon number is $\langle n \rangle = \abs{\alpha}^2$, and without loss of generality we choose the phase of the input light such that $\alpha = i\abs{\alpha}$, and thus $\langle X \rangle  = 0$ and  $\langle P \rangle = \abs{\alpha}$. Defining  $\varDelta \h{X} \equiv \hat{X} - \langle X \rangle$ and $\varDelta \h{P} \equiv \h{P} - \langle P \rangle$, the photon number operator can be written
\begin{align}
\nonumber    \h{n} & = 
\h{X}^2 + \h{P}^2 - \nonumber   \frac{1}{2}\\
\nonumber   & = \langle P \rangle^2 + 2 \langle P \rangle \varDelta \h{P} + \left( \varDelta \h{X}^2 + \varDelta \h{P}^2 - \frac{1}{2} \right) \\
            & \approx \abs{\alpha}^2 + 2\abs{\alpha}  \varDelta \h{P}.
\end{align}
We note that for a coherent state, the dropped term in parentheses has zero mean and  variance $1/4$. We neglect this term in the light-matter interaction Hamiltonian $\h{H}_{\rm qnd}$, considering that $|\alpha|^2 \gg 1$. Thus $\vtext{\tau} \h{H}_{\rm qnd}$ reads:
\begin{equation}
    \modtext{\tau} \h{H}_{\rm qnd} = g \left(\h{J_0} - \h{J_z}\right) \left(\abs{\alpha}^2 + 2 \abs{\alpha} \varDelta \h{P}\right).
\end{equation}

\noindent The term proportional to $\vert \alpha \vert^2$ is the ``classical'' contribution to the net rotation. We assume $g |\alpha|^2$ is an integer multiple of $2\pi$. Using the input-output Heisenberg formalism for small $\modtext{\tau}$, the net result of the pulse can then be written
\begin{eqnarray}
\label{eq:InOutJxNL}
    \h{J}_x(t+\tau)  &=& \h{J}_x(t)+ \hat{J}_y 2 g \rtnav  \varDelta \h{P}(t), \label{eq:line1}\\
\label{eq:InOutJyNL}
    \hat{J}_y(t+\tau)  &=& \hat{J}_y(t) - \h{J}_x  2 g \rtnav  \varDelta \h{P}(t), \label{eq:line2}\\
\label{eq:InOutDXNL}
    \varDelta \h{X}(t+\tau)  &=& \varDelta \h{X}(t) + g \rtnav  \left[\h{J_0}(t) -\h{J}_z(t) \right], \label{eq:line3}
\end{eqnarray}
\noindent along with the trivial relations $\h{J_0}(t+\tau) - \h{J_0}(t) = \h{J}_z(t+\tau) - \h{J}_z(t) = \h{P}(t+\tau) - \h{P}(t) = 0$. In the above relations, \autoref{eq:InOutJxNL} and \autoref{eq:InOutJyNL} describe a rotation of $\h{\mathbf{J}}$ about $z$ by  the small angle $\Delta \theta = 2 g |\alpha| \Delta \h{P}$, keeping terms only to first order in $\Delta \theta$.  Similarly, \autoref{eq:InOutDXNL} describes the result of a small optical phase shift  $g  (\h{J_0} -\h{J}_z )$.

\autoref{eq:InOutJxNL} and \autoref{eq:InOutJyNL} are nonlinear, due to the products $\h{J}_x \varDelta \h{P}$ and $\hat{J}_y \varDelta \h{P}$.  We note that during the Rabi protocol, $\h{J}_x$ and $\hat{J}_y$ evolve to take on large values $\sim N$, whereas their uncertainties are $\sim N^{1/2}$. Also  $\varDelta \h{P}$ remains small, of order unity, provided the optical phase shift remains small, as we assume.  This motivates writing $\h{J}_x(t) \varDelta \h{P} = [\langle J_x(t) \rangle + \varDelta \h{J}_x(t)] \varDelta \h{P} \approx \langle J_x(t) \rangle \varDelta \h{P}$ and similar for $\h{J}_y(t) \varDelta \h{P}$, at which point the nontrivial input-output relations become
\begin{eqnarray}
 &\h{J}_x(t+\tau) =   \h{J}_x(t) +  2 g \rtnav \langle J_y\rangle  \varDelta \h{P} \\
 &\hat{J}_y(t+\tau) = \hat{J}_y(t) -  2 g \rtnav \langle J_x\rangle  \varDelta \h{P} \\
 &\varDelta \h{X}(t+\tau) = \varDelta \h{X}(t) +  g \rtnav \h{J_0} - g \rtnav \h{J}_z.
\end{eqnarray}

Writing the phase-space vector $\h{\bm{V}} \equiv (\h{J_0}, \h{J}_x, \h{J}_y, \h{J}_z, \varDelta \h{X}, \varDelta \h{P})^T$, the input-output relations can be cast in matrix form $\h{\bm{V}}(t+\tau) = \mathcal{S}_{\rm qnd} \h{\bm{V}}(t)$, where 
\begin{eqnarray}
\mathcal{S}_{\rm qnd} = 
\begin{pmatrix}
1 & 0 &  0 & 0& 0& 0\\
0 & 1 &  {0} & 0& 0& 2 g \rtnav \langle  J_y \rangle\\
0 & {0} &  1 & 0& 0& -2 g \rtnav \langle J_x \rangle \\
0 & 0 &  0 & 1& 0& 0 \\
 g \rtnav & 0 &  0 & -  g \rtnav & 1& 0 \\
0 & 0 &  0 & 0& 0& 1 \\
\end{pmatrix}.\label{eq:qndM}
\end{eqnarray}

\subsection{Initial conditions}

The initial conditions used in our calculation for the three-measurement protocol and the reference protocol are summarized in \autoref{tab:initial_cond} below.

\begin{table}[h!]
    \centering
    \begin{tabular}{|c|c|c|c|}
    \hline
    Operator & $\langle \mathcal{O} \rangle$ & $(\Delta \mathcal{O})^2_{\rm ref}$ & $(\Delta \mathcal{O})^2_{\rm 3_{meas}} $ \\
    \hline
     $\h{J_0}$    & N & 0 & N/2 \\
     \hline
     $\h{J}_x$    & 0  & N/4 & N/4  \\
     \hline
     $\hat{J}_y$    & 0  &N/4 & N/4 \\
     \hline
     $\h{J}_z$    & -N & 0  & N/2 \\
     \hline
     $\varDelta \h{X}$ & 0 & 1/4 & 1/4 \\
     \hline
     $\varDelta \h{P}$ & 0 & 1/4 & 1/4 \\
     \hline
    \end{tabular}
    \caption{Initial conditions used for the calculations of the reference and three-measurement protocols, as described in the main text.}
    \label{tab:initial_cond}
\end{table}

\section{Model for light-atom coupling strength and experimental parameters}
In order to simulate realistic experimental situations, we use the parameters reported in recent literature \cite{vallet_noise-immune_2017}. The atom-light coupling strength is
\begin{equation}
    g = \frac{1}{S_{\rm eff}} \frac{3 \lambda^2 }{2 \pi} \frac{\Delta_{\rm qnd}/\Gamma}{s+4 (\Delta_{\rm qnd}/\Gamma)^2},
\end{equation}
where $s = P_c/(S_{\rm eff} I_{\rm sat})$ is the saturation parameter with $P_c$ the intra cavity power, and $S_{\rm eff} = \SI{2.9e3}{\micro\meter\squared}$ is an effective area taken from the same source. $\Gamma = 2 \pi \times 32$ MHz corresponds to the $^1S_0 - ^1P_1$ dipole transition in $^{87}$Sr, $\Delta_{\rm qnd} = 2 \pi \times 920$ MHz corresponds to the probing detuning reported, and the incoherent scattering rate is given by $\eta = \Gamma/2 \times  s/(s+4(\Delta_{\rm qnd}/\Gamma)^2)$.

\newcommand{\bmu}{\boldsymbol{\mu}}

\section{Covariance Matrix Calculations}

To compute $\bmu$ and $\Gamma_X$ we employ established covariance matrix methods \cite{MadsenPRA2004, KoschorreckJPAMOP2009, Colangelo2013,kraus_entanglement_2003,molmer_estimation_2004,PetersonPRA2006}. 
We describe the spin system and optical probe pulses with the phase-space vector $\h{\bm{V}} =  \h{\bm{J}} \oplus \left(\bigoplus_{i=1}^{N_{\rm pulses}} \varDelta \h{\quadvec}_i \right)$, where $\h{\bm{J}}=(\h{J}_0,\h{J}_x,\h{J}_y,\h{J}_z)^T$, $\h{\quadvec}_i \equiv ( \h{X}_i, \h{P}_i)^T$ describes the $i$th optical pulse. We assume that the state is Gaussian, and remains so during the entire measurement sequence.
The system is thus completely characterized by the vector of first moments $\langle \bm{V} \rangle$ and the covariance matrix $\Gamma_V = 1/2 ~ \langle \bm{V} \otimes \bm{V} + (\bm{V} \otimes \bm{V})^{T}\rangle - \langle \bm{V} \rangle \otimes \langle \bm{V} \rangle$.
$\langle \bm{V} \rangle$ and $\Gamma_V$ evolve deterministically through the sequence of coherent Rabi oscillations generated by $\h{H}_{\rm rot}$, and sudden
light-matter interactions generated by $\h{H}_{\rm qnd}$. As mentioned in the main text, the transformation of the state due to the action of $\hat{H}_{\rm qnd}$ is computed by integrating  $d\h{\mathbf{V}}/dt = {-i} [\h{\mathbf{V}},\h{H}_\mathrm{qnd}]$ and dropping terms beyond first order in quantum fluctuations, to find the linear relations \cite{Colangelo2013}
\begin{eqnarray}
    \h{\bm{V}} (t+\tau) &=& \mathcal{S}_{\rm qnd} \h{\bm{V}}(t)\\
    \Gamma_V (t+\tau) &=& \mathcal{S}_{\rm qnd}  \Gamma_V(t) \mathcal{S}_{\rm qnd}^{T},
\end{eqnarray}
where the explicit form of $\mathcal{S}_{\rm qnd}$ for a single QND measurement is given in \autoref{eq:qndM}.

We include the effect of loss and decoherence due to atom-photon scattering during the probing via
\begin{eqnarray}
    \h{\bm{V}}'(t + \tau)  &=& \mathcal{M} \h{\bm{V}}(t + \tau)\\
    \Gamma'_V(t + \tau) &=& \mathcal{M} \Gamma_V(t + \tau) \mathcal{M}^T + \mathcal{N},
\end{eqnarray}\label{eq:decoherence}
where $\mathcal{M} = [(1-\eta) \mathbbm{1}_{4}] \oplus \mathbbm{1}_{2}$ , and $\mathcal{N} = [\eta (1-\eta)  \langle N \rangle \Gamma_{\lambda} +  \langle N \rangle \eta \frac{f (f+1)}{3} \mathbbm{1}_4 ] \oplus [\mathbbm{O}_2]$.
Here $\eta = \eta_{\gamma} \abs{\alpha}^2$ is the fraction of atoms that scatter a photon due to the interaction with a probe pulse containing $\abs{\alpha}^2$ photons with incoherent scattering rate $\eta_\gamma$, and $\Gamma_{\lambda}$ is the single atom covariance matrix of the atomic state \cite{Colangelo2013}. Multi-pulse sequences are constructed analogously \cite{KoschorreckJPAMOP2009, Colangelo2013}. The model gives unsightly but useful analytic results for $\bmu$ and $\Gamma_X$, which can be read off directly from $\langle \bm{V} \rangle$ and $\Gamma_V$.

\section{Derivation of the three-measurement protocol sensitivity}

As described in the main text, $\textbf{X} \equiv \{ X_l \}$ is the column random vector of measurement outcomes of the post-interaction optical variable $\h{X}^{(out)}$, $\boldsymbol{\mu} = \langle \textbf{X}(\delta) \rangle$ is the vector containing the expected values of $\bf{X}$, and $\Gamma_X$ is the covariance matrix of these measurements.
We assume that the measurement outcomes are random variables described by a multivariate Gaussian probability distribution:

\begin{equation}
p(\textbf{X}|\delta) = \frac{1}{\sqrt{(2 \pi)^d|\Gamma_X|} } \exp[ -\frac{1}{2}(\textbf{X}-\boldsymbol{\mu})^T \Gamma_X^{-1} (\textbf{X}-\boldsymbol{\mu})]. \label{eq:pdf}
\end{equation}

In our problem, these quantities are calculated using the covariance matrix formalism. Following the standard approach, we seek to minimize the quantity on the exponential of equation \autoref{eq:pdf}, 
\begin{equation}
    W \equiv \frac{1}{2}(\textbf{X}-\boldsymbol{\mu})^T \Gamma_X^{-1} (\textbf{X}-\boldsymbol{\mu}).
\end{equation}
We linearly approximate around the nominal value of the detuning $\delta_0$, i.e., we make the approximation $\boldsymbol{\mu}(\delta) \approx \boldsymbol{\mu}_0 + \boldsymbol{\mu}^{\prime}~\Delta\delta$, where $\boldsymbol{\mu}_0 = \boldsymbol{\mu}(\delta_0)$, $\boldsymbol{\mu}^{\prime} = \partial_{\delta}\boldsymbol{\mu}(\delta)|_{\delta_0}$, and $\Delta \delta = \delta - \delta_0$. The minimum of $W$ will then satisfy
\begin{equation}
\partial_{\delta} W \propto (\boldsymbol{\mu}^{\prime})^T \Gamma_X^{-1} (\textbf{X} - \boldsymbol{\mu}_0 - \boldsymbol{\mu}^{\prime} ~ \Delta \delta) = 0
\end{equation}
which is solved by 
\begin{equation}
\Delta \delta = \frac{(\boldsymbol{\mu}^{\prime})^T \Gamma_X^{-1} (\textbf{X} - \boldsymbol{\mu}_0)}{(\boldsymbol{\mu}^{\prime})^T \Gamma_X^{-1} \boldsymbol{\mu}^{\prime}}. \label{eq:detuningerr}
\end{equation}
Note that in deriving this expression, we assumed that $\partial_\delta \Gamma_X^{-1}$ is of the order of $\Delta \delta^2$, which is reasonable under the linear approximation. From \autoref{eq:detuningerr}, we can derive an estimator $\estimator{\delta}$ for the clock laser detuning $\delta$:
\begin{equation}
\estimator{\delta} = \delta_0 + \frac{(\boldsymbol{\mu}^{\prime})^T \Gamma_X^{-1} (\textbf{X} - \boldsymbol{\mu}_0)}{(\boldsymbol{\mu}^{\prime})^T \Gamma_X^{-1} \boldsymbol{\mu}^{\prime}}. 
\end{equation}\label{eq:estimator}

\noindent Correspondingly, the sensitivity of the protocol can be calculated from the expected mean squared error of this estimator
\begin{equation}
    E[(\estimator{\delta}-\delta)^2] = \int d \textbf{X} ~ p(\textbf{X}|\delta)~ (\Delta\delta)^2. \label{eq:estimator1}
\end{equation}

\noindent To evaluate the integral in \autoref{eq:estimator1} we note that the covariance matrix $\Gamma_X$ is a real, symmetric matrix. Thus, we can always find an orthonormal basis $v_i$ in which $\Gamma_X = \lambda_i (v_i \wedge v_i)$ where summation over repeated indices is implicit. In this basis, we call $X_i = v_i^{T} \textbf{X}$ the $ith$ component of the vector $\textbf{X}$. Then, we can write  $(\Gamma_X)_{ij} = \delta_{ij} \lambda_i$, and its determinant $|\Gamma_X| = \prod_i \lambda_i$. In this basis $p(\textbf{X}|\delta)$ factorizes, and 

\begin{equation}
    (\Delta \delta)^2 =  \frac{\mu^{\prime}_i \lambda_i^{-1} (X - \mu_0)_i ~ \mu^{\prime}_j \lambda_j^{-1} (X - \mu_0)_j }{(\mu^{\prime}_i \lambda_i^{-1} \mu^{\prime}_i)^2}.
\end{equation}

\noindent From the expression above and the previous considerations, we can observe that $E[(X-\mu_0)_i (X-\mu_0)_j ] = E[(X-\mu)_i] E[(X-\mu)_j] = 0$ for $i \neq j$, and $E[(X-\mu)_i (X-\mu)_i ] = E[(X-\mu)^2_i] = \lambda_i$. Therefore, the expected mean squared error:

\begin{equation}
    E[(\Delta \delta)^2] = \frac{(\mu^{\prime}_i \lambda_i^{-1})^2 \lambda_i}{((\mu^{\prime})_i \lambda_i^{-1} \mu^{\prime}_i)^2} = \frac{1}{(\boldsymbol{\mu}^{\prime})^T \Gamma_X^{-1} \boldsymbol{\mu}^{\prime}}
\end{equation}

\section{Spin squeezing witness}

If  $\{\h{J}_j,\h{J}_k,\h{J}_l\}$ are orthogonal components of $\h{\mathbf{J}}$, all non-entangled states, including those with uncertain particle number \cite{hyllus_entanglement_2012}, obey
\begin{eqnarray}
{(\Delta J_l)^2} \ge \langle\frac{ \h{J}_j^2}{\h{N}-1} \rangle +  \langle\frac{ \h{J}_k^2}{\h{N}-1} \rangle +  \frac{1}{2} \langle\frac{\h{N}}{\h{N}-1} \rangle. 
\label{eq:SpinSqueezingInequality}
\end{eqnarray} 

In the large-$N$ scenario that concerns us, this becomes
\begin{equation}
     (\Delta J_l)^2 \ge \langle \frac{\h{J}_j^2}{\h{N}} \rangle + \langle \frac{\h{J}_k^2}{\h{N}} \rangle.
    \label{eq:witness}
\end{equation} 

\noindent From the covariance matrix calculation, we don't have access to $\langle \h{J}_{k,l}^2/\h{N}\rangle$. Nevertheless, within the Gaussian approximation we can approximate $\langle \h{J}_{k}^2/\h{N}\rangle$ by Taylor expansion about $\langle J_k \rangle^2/\langle N \rangle$:
\begin{eqnarray}
\frac{\h{J}_k^2}{\h{N}} &=& 
\frac{\langle J_k\rangle^2}{\langle N \rangle } + 2\frac{\langle J_k\rangle}{\langle N \rangle } \varDelta \h{J}_k
- \frac{\langle J_k\rangle^2}{\langle N \rangle^2 } \varDelta \h{N}
\nonumber \\ & & + 
\frac{ (\varDelta \h{J}_k)^2}{\langle N \rangle}
- 2\frac{\langle J_k\rangle}{\langle N \rangle^2 } \varDelta \h{J}_k \varDelta \h{N}
+ 2\frac{\langle J_k\rangle^2}{\langle N \rangle^3} (\varDelta \h{N})^2 
\nonumber \\ & & +  O(\varDelta \h{N}, \varDelta \h{J}_k)^3,
\end{eqnarray}
where $\varDelta \h{A} \equiv \h{A} - \langle A \rangle$ denotes small deviations around the mean. When we calculate the expectation value of the latter expression, the first-order terms vanish, the third-order terms can be neglected in the Gaussian approximation, and we are left with 
\begin{eqnarray}
\langle \frac{\h{J}_k^2}{\h{N}} \rangle &\approx& 
\frac{\langle J_k\rangle^2}{\langle N \rangle}  + 
\frac{\Delta J_k^2}{\langle N \rangle}
- 2\frac{\langle J_k\rangle}{\langle N \rangle^2 } \Cov[J_k,N]
+ 2\frac{\langle J_k\rangle^2}{\langle N \rangle^3} \Delta N^2
\nonumber\\ & = & 
\frac{\langle J_k\rangle^2}{\langle N \rangle } \left( 1 + 
\frac{\Delta J_k^2}{\langle J_k \rangle^2}
- 2 \frac{\Cov[J_k,N]}{\langle J_k\rangle \langle N \rangle} 
+ \frac{\Delta N^2}{\langle N \rangle^2}  \right),
\end{eqnarray}
where $(\Delta A)^2 \equiv \langle A^2 \rangle - \langle A \rangle^2$ denotes the variance of the operator. The above terms can be computed using the mean and covariance matrix elements.

\bibliographystyle{apsrev4-1no-url}
\bibliography{OLC_dbo_suppl}